\documentclass[a4paper,11pt]{article}
\pdfoutput=1 

\usepackage{jcappub} 

\usepackage{graphicx}	% Including figure files
\usepackage{hyperref}   % hyperref package all cross-referenced elements become hyperlinked.
\usepackage{amsmath}    % Advanced maths commands

\title{Redshift effects implications on revised models of Stephan's quintet}

\author[a]{M. L. Abdelali$^1$\note{Corresponding author.}}
\author[a]{and N. Mebarki}

\affiliation[a]{Laboratoire de Physique Mathematique et Subatomique, Universite Freres Mentouri Constantine 1, BP 325 Route de Ain El Bey, Constantine 25017, Algeria}

\emailAdd{mabdelali1@gmail.com}
\emailAdd{nnmebarki@yahoo.fr}

\abstract{Recent observations of Stephan's Quintet gave new indications on its formation scenario. 
Older formation and a role of NCG 7317 should be considered in revised numerical models of the compact group. 
Velocities of group members to recreate are estimated from redshift measurements. 
Several effects contribute to observed redshifts and a new effect is predicted to be the result of the gravitational interaction between photons and constant magnetic fields creating gravitational waves. 
The energy carried by these waves is manifested as redshifts of photons. 
Cosmological simulations data are used to prove the significant contribution of our effect. 
The analysis of synthetic observations created from those simulations has shown that redshifts differences of SQ members could be misinterpreted as caused only from Doppler Effect. 
The revised models of the group should consider a new method to recreate the formation scenario based on redshift patterns and not mis-estimated velocities. 
}


\begin{document}
\maketitle
\flushbottom

\section{Introduction}

The famous Stephen's Quintet (SQ) was subject of several previous studies (e.g. see ref. \cite{RId149} and \cite{RId148}). 
This compact group of gravitationally interacting galaxies is a place of interesting phenomena and a test field for various theories. 
Arp (1973)\cite{NRId010} believed that the light from NCG 7320 is affected by non-Doppler effects making it appear to be a foreground galaxy. 
Later observations indicated the role of NCG 7320c in the formation scenario of SQ and gave evidence to the foreground position of NCG 7320. 
Numerical simulations were done to test models recreating the evolution history of this cluster (e.g. see Renaud, Appleton \& Xu (2010)\cite{RId153} hereafter RAX10). 
Recently, Duc, Cuillandre \& Renaud (2018)\cite{NRId001} (hereafter DCR18) revisited SQ and observed diffuse, reddish, lopsided, halo extended toward NGC 7317. 
This halo made of old stars indicates a group formation several Gyr ago earlier than previously thought and a role of NCG 7317 in the early construction of the group. 
Those observations change our perception of the history of SQ and revised models and scenarios should be studied. 
Our current models use previous data and indications to reproduce observed features of the group including velocities of each SQ member and its morphology. 
These radial velocities are estimated from redshift measurements and are very influential on the model evolution. 

Redshift is caused by several effects: Cosmological redshift $Z_C$, Doppler redshift $Z_D$ and Gravitational redshift $Z_G$. 
The observed redshift $Z_{Obs}$ represents a combination of all those effects and the measured value is giving by
\begin{equation}
(1+Z_{Obs})=(1+Z_C)(1+Z_D)(1+Z_G)(1+Z_{NC}),
\end{equation}
where $Z_{NC}$ stands for any other kind of redshift effects (of non cosmological nature) that could affect the travelling photons. 
Along with cosmological redshift, Doppler Effect impacts are estimated by considering the peculiar motion of galaxies. 
The gravitational redshift contribution was considered to be negligible. 
A number of recent articles had investigated the cosmological parameters bias caused by gravitational redshifts (e.g. ref. \cite{RId131, RId133}). 
In the study of Wojtak, Davis \& Wiis (2015) (see ref. \cite{RId003} hereafter WDW15), they used numerical simulations data to estimate the resulting gravitational redshift contribution to observations and found $2\times10^{-5}$ additional shift influencing estimations of cosmological parameters. 
A recent work of Calcino \& Davis (2017) \cite{RId168} had used new statistical methods to evaluate the redshift bias $\Delta z$ and estimated this shift to $2.6\times10^{-4}$. 
Their explanation was that possibly gravitational redshifts have more contribution or that measurement errors were responsible of the difference of bias estimations. 
As an interesting possibility, this additional estimated bias may represent the contribution of another redshift effect other than cosmological, Doppler or gravitational redshifts. 
This contribution must not be blamed on measurement errors. 

In this article, we present new redshift effect affecting propagating photons within magnetic fields. 
Numerical simulations are argued to be the best way to estimate the contribution in cosmological and astrophysical data. 
In those simulations, the synthetic observations of a compact group similar to SQ are giving new indications that could alter our perception to formation scenarios of the quintet. 
The article is organized as follows: In section \ref{sec:3}, we introduce our new redshift effect with its mathematical formalism and discuss its significant contribution to cosmological and astrophysical data. 
In sections \ref{sec:6}, we present our approach to evaluate our redshift effect impacts. 
In sections \ref{sec:7}, we present cosmological simulations done and the algorithm for the creation of synthetic observations. 
In sections \ref{sec:8}, we discuss our results and their implications on revised SQ models. 
We finish the article by main conclusions in section \ref{sec:9}.

\section{New magnetically induced redshift: mathematical formalism}
\label{sec:3}

The mechanism of our redshift effect starts with photons travelling through external (or background) magnetic fields. 
The gravitational interaction of electromagnetic waves with these external static magnetic fields is studied using Einstein general relativity equations (see equation 8.14 in ref. \cite{NRId011})
\begin{equation}
R_{\mu \nu} - \frac{1}{2} g_{\mu \nu} R = - \kappa  T_{\mu \nu}, 
\end{equation}
where $\kappa = \frac{8\pi G}{c^4}$ with $G$ is the gravitational constant and $c$ the speed of light. 
$R$ is the curvature scalar computed from the Ricci tensor which is $R_{\alpha\beta} = R^{\gamma}_{\alpha\beta\gamma}$ and $R^{\rho}_{\alpha\beta\gamma}$ is Riemann tensor. 
The left hand side represents the Einstein tensor computed from the space-time metric $g_{\mu \nu}$. 
The right hand side represents the energy momentum tensor of the studied source of space-time curvature. 
The energy of electromagnetic fields is weak, but still curves space-time. 
Then, the linear approximation of Einstein equations is used starting from the metric that has a weak perturbation $h_{\mu \nu}$ to Minkowski metric $\eta_{\mu \nu}$ which takes then the following form 
\begin{equation}
g_{\mu \nu} = \eta_{\mu \nu} + h_{\mu \nu}.
\end{equation}
In the first order of the approximation, Einstein equations are expressed in term of this weak metric perturbation as
\begin{equation}
 {\partial}_{\alpha}{\partial}^{\alpha} \overline{h}_{\mu\nu} + 
  \eta_{\mu \nu} {\partial}_{\alpha}{\partial}_{\beta} \overline{h}^{\alpha\beta}-
  {\partial}_{\nu}{\partial}_{\alpha} \overline{h}^{\alpha}_{\mu}-
 {\partial}_{\mu}{\partial}_{\alpha} \overline{h}^{\alpha}_{\nu}
 = -2 \kappa  T_{\mu \nu}, 
\end{equation}
where
\begin{equation}
\overline{h}_{\mu\nu} = h_{\mu\nu} - \frac{1}{2} \eta_{\mu \nu} h^{\alpha}_{\alpha}.
\end{equation}
We choose to consider solutions verifying the transverse-traceless gauge which simplifies Einstein equations even further to
\begin{equation}
{\partial}_{\alpha}{\partial}^{\alpha} \overline{h}_{\mu\nu} = -2 \kappa  T_{\mu \nu},
\label{eq:TTEeqs}
\end{equation}
where ${\partial}_{\beta} {\overline{h}}^{\alpha\beta}=0$ , $h^{\alpha}_{\alpha}=0$ and ${\overline{h}}_{\mu\nu}={h}_{\mu\nu}$. 
These differential equations are wave equations with source function that has a general solution and special solution. 
The general solution is just the usual gravitational waves with their two polarisations. 
We are interested in the special solution related to our studied source which is the electromagnetic fields in our case. 
After the special solution of these equations is found, we check if these gravitational waves described by this metric perturbation are physical radiations carrying energy. 
The second order Einstein tensor represents the energy momentum tensor of emitted gravitational waves (see ref. \cite{NRId011}) which could be written as
\begin{equation}
t_{\mu\nu}(h_{\rho\sigma}) = \frac{1}{\kappa}  \displaystyle{\langle {(R_{\mu \nu} - \frac{1}{2} g_{\mu \nu} R)}^{(2)} \rangle},
\end{equation}
where $\displaystyle{\langle \dots \rangle}$ denotes an average over a small region at each point in space-time and ${}^{(2)}$ denotes the second order in the perturbation expansion. 
In the transverse-traceless gauge, this energy-momentum tensor in vacuo is reduced to
\begin{equation}
t_{\mu\nu}(h_{\rho\sigma}) = \frac{1}{4 \kappa}  \displaystyle{\langle (\partial _{\mu} h_{\rho \sigma}) \partial _{\nu} h^{\rho \sigma} \rangle}.
\end{equation}
Thus, any physical gravitational radiation should have non vanishing energy momentum tensor $t_{\mu\nu}$. 

The energy momentum tensor is given for electromagnetic fields expressed in function of the electromagnetic field tensor $F_{\mu\nu}$ as
\begin{equation}
T^{(EM)}_{\mu \nu} = - \frac{1}{{\mu}_0} \left( F_{\mu\alpha} F_{\nu\beta} g^{\alpha\beta} - \frac{1}{4} g_{\mu\nu} F_{\alpha\beta} F^{\alpha\beta} \right),
\end{equation}
where
\begin{equation}
F^{\mu\nu} = 
\left(
\begin{array}{cccc}
0 & -E^1/c & -E^2/c & -E^3/c \\
E^1/c & 0 & -B^3 & B^2 \\
E^2/c & B^3 & 0 & -B^1 \\
E^3/c & -B^2 & B^1 &0 \\
\end{array}
\right),
\end{equation}
$\mu _0$ is the permeability of free space, $B^i\,i=1..3$ are magnetic fields components and $E^i\,i=1..3$ are electric fields components. 
The electromagnetic fields are composed of a plane monochromatic electromagnetic wave given by its electric field $\overrightarrow{E}$ ($E_0 \cos (k(t-z))$, $0$, $0$) and magnetic field $\overrightarrow{B}$ ($0$, $\frac{E_0}{c} \cos (k(t-z))$, $0$). 
The electromagnetic wave is propagating along the z-axis. 
We study the propagation of this electromagnetic wave in 3 regions of space depending on the existing constant magnetic field $\overrightarrow{B}_{ext}$ as follows: the first ($z < 0$) and the last region ($z > l$)) have no external magnetic fields only the second region ($0 < z < l$) has constant magnetic field, where $l$ represent the coherent length where the constant magnetic field exists. 
The magnetic fields are perpendicular to the direction of propagation along x-axis $\overrightarrow{B}_{ext}$ ($B_x$, $0$, $0$). 
The electromagnetic tensor of these fields in this second region of space is given by
\begin{equation}
\begin{split}
F^{\mu \nu} = \frac{E_0 \cos (k(t-z))}{c} 
\left(
\begin{array}{cccc}
0 & -1 & 0 & 0 \\
1 & 0 & 0 & 1 \\
0 & 0 & 0 & 0 \\
0 & -1 & 0 &0 \\
\end{array}
\right) \\
+B_x
\left(
\begin{array}{cccc}
0 & 0 & 0 & 0 \\
0 & 0 & 0 & 0 \\
0 & 0 & 0 & -1 \\
0 & 0 & 1 & 0 \\
\end{array}
\right).
\end{split}
\end{equation}
For this second region, we can distinguish three parts of the electromagnetic energy momentum tensor: the first is static related only to constant magnetic fields, second part is related only to the electromagnetic wave and the third part is proportional to both magnetic fields and electromagnetic wave. 
This tensor has the following form
\begin{equation}
\begin{split}
T^{(EM)}_{\mu \nu} = \frac{B^2_x}{2{\mu}_0}
\left(
\begin{array}{cccc}
1 & 0 & 0 & 0 \\
0 & -1 & 0 & 0 \\
0 & 0 & 1 & 0 \\
0 & 0 & 0 & 1 \\
\end{array}
\right) \\
+\frac{E^2_0 {\cos}^2 (k(t-z))}{{\mu}_0 c^2} 
\left(
\begin{array}{cccc}
1 & 0 & 0 & -1 \\
0 & 0 & 0 & 0 \\
0 & 0 & 0 & 0 \\
-1 & 0 & 0 & 1 \\
\end{array}
\right) \\
+\frac{B_x E_0 \cos (k(t-z))}{{\mu}_0 c}
\left(
\begin{array}{cccc}
0 & 0 & 0 & 0 \\
0 & 0 & -1 & 0 \\
0 & -1 & 0 & 0 \\
0 & 0 & 0 & 0 \\
\end{array}
\right).
\end{split}
\end{equation} \\
In the first and third regions, only the terms related to electromagnetic wave is not vanishing. 
After replacing this tensor in \eqref{eq:TTEeqs}, we solve the gravitational equations to find the perturbation metric. 
We have rewritten the perturbation metric to three parts according to the three parts of $T^{(EM)}_{\mu \nu}$ as follows 
\begin{equation}
\begin{split}
\overline{h}_{\mu\nu} = \overline{h}^{(I)}_{\mu\nu} + \overline{h}^{(II)}_{\mu\nu} + \overline{h}^{(III)}_{\mu\nu} \\
=
f^{(I)}(t,z)
\left(
\begin{array}{cccc}
1 & 0 & 0 & 0 \\
0 & -1 & 0 & 0 \\
0 & 0 & 1 & 0 \\
0 & 0 & 0 & 1 \\
\end{array}
\right) \\
+ f^{(II)}(t,z)
\left(
\begin{array}{cccc}
1 & 0 & 0 & -1 \\
0 & 0 & 0 & 0 \\
0 & 0 & 0 & 0 \\
-1 & 0 & 0 & 1 \\
\end{array}
\right) \\
+ f^{(III)}(t,z)
\left(
\begin{array}{cccc}
0 & 0 & 0 & 0 \\
0 & 0 & -1 & 0 \\
0 & -1 & 0 & 0 \\
0 & 0 & 0 & 0 \\
\end{array}
\right).
\end{split}
\end{equation} \\
This reduces our equations to three simple wave equations with source terms. 
The general solution to these equations represents the ordinary gravitational waves propagating in vacuum. 
We are interested to the special solutions related to our electromagnetic fields. 
After a straightforward operation, we find that the two first parts produces non physical gravitational radiations with vanishing energy momentum tensor and violate the traceless transverse gauge. 
These two parts are related to pure magnetic field and pure electromagnetic wave contributions. 
This is the same situation in first and last region where only pure electromagnetic wave contributions exists. 
We find that the physical radiations ($h^{(III)}_{\mu\nu}$) given by 
\begin{equation}
h^{(III)}_{\mu\nu} = \frac{-2\kappa E_0 B_x}{\mu _0 c} \frac{z}{2k}\sin (k(t-z))
\left(
\begin{array}{cccc}
0 & 0 & 0 & 0 \\
0 & 0 & -1 & 0 \\
0 & -1 & 0 & 0 \\
0 & 0 & 0 & 0 \\
\end{array}
\right),
\end{equation} \\
has a non vanishing energy momentum tensor of and doesn't violate the traceless transverse gauge. 
It is related to cross terms between constant magnetic fields and the propagating electromagnetic wave. 
The gravitational radiation created by this term has one of the two known polarization of gravitational waves. 
This solution verify the continuity conditions at the boundaries. 
First between region 1 and 2 at $z=0$, $h^{(III)}_{\mu\nu}$ vanishes indicating that these gravitational waves are generated only from the interaction of incident electromagnetic waves with the magnetic fields background. 
This process doesn't require an initial or incident gravitational waves to occur which are not hypothesized in our scenario in the first place. 
Between region 2 and 3 at $z=l$, $h^{(III)}_{\mu\nu}$ start to have a constant amplitude for these gravitational waves. 
These waves propagate then as any other ordinary gravitational waves in vacuum. 
We have the non vanishing elements of energy momentum tensor for the radiated gravitational waves as follows
\begin{equation}
t_{00} = \frac{\kappa}{4 {\mu}^2_0 c^2} B^2_x E^2_0 z^2,
\end{equation}
\begin{equation}
t_{33} = \frac{\kappa}{4 {\mu}^2_0 c^2} \frac{B^2_x E^2_0}{k^2} (k^2 z^2 +1)
\end{equation}
and
\begin{equation}
t_{03} = -\frac{\kappa}{4 {\mu}^2_0 c^2} B^2_x E^2_0 z^2.
\label{eq:GWEnergy}
\end{equation}

Thus, gravitational waves created by electromagnetic waves in vacuum without the existence of magnetic fields have null energy momentum tensor. 
This is the case in first and last regions. 
Only within the second region with magnetic fields, the electromagnetic wave creates a physical gravitational radiation carrying energy. 
We expect an energy loss of the radiating system which is photons in our case. 
In the case of pulsar binary, we observed a loss in rotation period when radiating gravitational waves and we observe a spin-up of binary systems. 
In analogy, the radiation of gravitational waves makes the electromagnetic wave lose energy expressed as frequency drop or redshift. 
The radiated energy carried by gravitational waves is calculated from their energy momentum tensor components (see ref. \cite{NRId011}) as follows
\begin{equation}
\frac{dE}{dt} = -L_{GW} = -\int\int F(\overrightarrow{n})dS,
\end{equation}
where the right hand side contains the gravitational waves energy flux. 
It is calculated from the energy momentum tensor as follows
\begin{equation}
F(\overrightarrow{n}) = -c t^{0k} n_k,
\end{equation}
where
\begin{equation}
[n_{\mu}] = [1, 0, 0, -1].
\end{equation}
When replacing the tensor elements, we find
\begin{equation}
\frac{dE}{cdt} = -\int\int \frac{\kappa}{4 {\mu}^2_0 c^2} B^2_x E^2_0 z^2 dx dy,
\label{eq:energyloss}
\end{equation}
where $E$ represent the energy of the photon. 
Poynting vector represent the power density vector associated with an electromagnetic field. 
The time average over the oscillation period of Poynting vector is called the flux density, irradiance or intensity of light wave $I$ (see ref. \cite{NRId012}) which is given for a plane monochromatic light wave as
\begin{equation}
I = \frac{E^2_0}{2\mu _0 c}.
\end{equation}
In another hand, the energy of light is carried by discrete particles called photons in the perspective of quantum mechanics. 
If the light has a frequency of $\nu$, then the photon's energy is $h \nu$. 
The intensity of the light is equal to the number of photons $F$ crossing a unit area, in a unit time, multiplied by the energy of an individual photon 
\begin{equation}
I = F h \nu.
\end{equation}
When considering in our case of a light wave propagating in z-axis, it could be seen as a beam of single photons travelling only along z-axis. 
We introduce a delta function in $F$ to describe this propagation mathematically: $\delta(x)$ impose the photon to exist only in $x = 0$ and $\delta(y)$ impose the photon to exist only in $y = 0$ representing a propagation of photon particle along z-axis. 
Then, the number of photons $F$ is 
\begin{equation}
F=\delta(x)\delta(y)\frac{photon}{m^2 s}.
\end{equation}
From the last three equations, we could then put
\begin{equation}
\frac{E^2_0}{2\mu _0 c} = h \nu \delta(x)\delta(y)
\end{equation}
when we replace each term and simplify \eqref{eq:energyloss}, we find the following differential equation, after replacing $cdt=dz$ as the propagation is along the z-axis
\begin{equation}
\frac{d\nu}{dz}= \frac{-\kappa}{2 {\mu}_0 c} B^2_x \nu z^2 \int\int \delta(x)\delta(y) dxdy
\label{eq:zMIR00}
\end{equation}
that we integrate to find
\begin{equation}
\ln \left( \frac{\nu}{{\nu}_0} \right) = \frac{-\kappa}{6 {\mu}_0 c} B^2_x l^3
\label{eq:zMIR0}
\end{equation}
and
\begin{equation}
(1+z_{MIR})= \left( \frac{{\nu}_0}{\nu} \right) = \exp \left(  \frac{\kappa}{6 {\mu}_0 c} B^2_x l^3 \right),
\label{eq:zMIR}
\end{equation}
where $\nu _0$ is the initial frequency of the photon, $\nu$ is the frequency of the photon when leaving the second region and $l$ is the coherent length of the magnetic field where the magnetic field is constant (represent here the length of the second region along $z$ axis).
This final result $z_{MIR}$ represents the redshift that a photon suffer when propagating within constant magnetic field and radiating gravitational waves. 
It could appear that the right hand side of equation \eqref{eq:zMIR0} has the dimension of time (seconds in SI units). 
But, it is worth to notice that a dimension of $s^{-1}$ remains after replacing the number of photons $F$ and simplifying its dimensions in equation \eqref{eq:zMIR00}. 
This ensures that the right hand side of equation \eqref{eq:zMIR0} is dimensionless. 

\section{New magnetically induced redshift: approach to impacts study}
\label{sec:6}

For the steps leading to radiated gravitational waves energy-momentum tensor (see \eqref{eq:GWEnergy}), our results are not different from some previous papers (see ref. \cite{RId042}) but with different interpretation. 
In a theorized phenomenon known as Gertsenshtein effect, light passing through an external magnetic field produces a gravitational wave via wave resonance. 
This idea was predicted in the pioneer papers of Gertsenshtein (1962) (see ref. \cite{RId039}) and Zeldovich (see ref. \cite{RId040}) with applications to some astrophysical phenomena. 
Gertsenshtein effect has some weak points in its conception and especially its interpretation of the final state. 
The conversion in quantum level of a spin 2 particle to a spin 1 particle could have a spin violation. 
The energy momentum tensor of the photon loses energy randomly in this effect indicating a non continuum and a non conservation of energy. 
Such effect has non observational data supporting it until now and no analogue effects to compare with it. 
There is no theoretical foundations that justify the interpretation of the energy momentum tensor elements of gravitational waves as probabilities of conversion. 
And there is no effect that made such interpretation and has experimental or observational evidence. 
In our effect, those gravitational waves carries energy radiated from the electromagnetic wave resulting in a redshift for the photon. 
The new effect has mainly two strong foundations that differentiate it from the Gertsenshtein effect. 
First, we have a continuity of the energy momentum tensor of both radiations. 
Second, this effect is built in analogy with binary system prediction of period (frequency) drop that was confirmed by Hulse-Taylor observations (e.g.\cite{RId169, RId170, RId171}). 
These reasons make our effect more coherent. 

We investigate the existence in cosmic conditions and significance. 
The existence of cosmic magnetic fields (CMF) has more and more observational evidence (see ref. \cite{RId059, RId058}). 
Magnetic fields in our solar system or in interstellar medium are strong enough but don't have large space for the photons to accumulate noticeable redshifts from our effect. 
But, at large scales of extragalactic and galactic mediums, magnetic fields are weak with ranges from $10^{-9}$ to $10^{-6} G$ but spread at large distances and have noticeable impacts on cosmological data. 
A magnetic field with strength of $10^{-6}G$ and a coherent length of $100kpc$ produces a $Z_{MIR}$ redshift of $2.71\times 10^{-2}$. 
We can conclude that only astrophysical and cosmological scales are the ones to have noticeable contributions from our effect. 
Cosmic magnetic fields constant over coherent lengths of $1Mpc$ have a weaker strength in order of $nG$ as shown by observations. 
Then, accumulation of our redshift effect should have significant but not inexplicably and unobserved high contributions to the total redshift of a distant extragalactic object. 
Observed redshifts are hardly (if not impossible in some cases) broken to their individual effect contributions especially if those contributions are in comparable ranges. 
Then, the contribution of our effect is mis-interpreted as caused by Doppler Effect. 
This affects current models recreating the kinematic evolution of local and global structure formations. 
The wrong interpretation of our effect contributions as gravitational redshifts would result in over-estimation of gravitational potentials and then of dark matter composition of clusters and the whole universe. 
Considering our effect contributions as cosmological redshifts is creating bias in cosmological parameters estimations and could account for additional estimations found by \cite{RId168}. 

Large scale surveys of extragalactic radio sources have important but not clear potential to investigate intergalactic magnetic fields. 
Techniques to discriminate Faraday rotations contributions are very ambiguous about the estimations degeneracy and possibility of over-estimations of each part (see ref. \cite{RId073, RId072}). 
Several limitations make the creation of a reliable and detailed all sky map a very challenging or impossible task. 
The estimation of our redshift effect impact on observations is not then feasible using current observational data on cosmic magnetic fields. 
Magneto-hydrodynamic simulations are taking more interest lately as they present new insights about the multi-variant and non-linear astrophysical questions. 
We need to use numerical simulations data to make impacts estimations of our new magnetically induced redshift. 
Some of available magneto-hydrodynamic simulations are limited in access and their outputs are not configured in the appropriate setting to our study. 
To see the bias and deviations of cosmological parameters, we need to create a complete synthetic observation: cosmological distances and redshifts and Doppler, gravitational and the new magnetically induced redshift effect. 

It is a main difference to previous studies done with simulation outputs (e.g. WDW15) where they didn't consider cosmological redshift. 
To make this possible, outputs of the simulations must be in specific redshifts intervals. 
Each dataset represents the state of the universe in the time (redshift) interval between two data dumps. 
The changes in such short redshift interval are negligible. 
With minimized intervals, the approximation is more accurate. 
In previous studies, observations were made from a single dataset taken in a fixed redshift - time (mostly present time $z=0.0$). 
After determining the set of output datasets needed to construct the appropriate observation, our customized simulations are executed.
%%%%%%%%%%%%%%%%%%%%%%%%%%%%%%%%%%%%%%%%%%%%%%%%%%%%%%%%%%%%%%%%%%%%%%%%%%%%%%%%%%%%%%%%%%%%%%%%%%%%%%%%%%%%%%%%%%%%%%%%%%%%%%%%%%%%%%%%%%%%%%%%%%%%%%%%%%

\section{Cosmological simulations and synthetic observations}
\label{sec:7}

Our simulation was performed using AMR code \texttt{ENZO} (see ref. \cite{RId119}) \footnote{http://enzo-project.org}. 
\texttt{ENZO} is a publically available code developed by an active community. 
It is a highly parallel code that was used for multi-physics cosmological magneto-hydrodynamics (MHD) simulations. 
It uses a particle-mesh N-body method to follow the dynamics of the Dark Matter and a variety of Riemann solvers to evolve the gas component.  
In our simulation, we fixed cosmological parameters to these values ($\Omega_b=0.04, \Omega_m=0.27, \Omega_\Lambda=0.73, H_0=0.71$), $\sigma _8$ to $0.9$ and the initial redshift to $z=99$. 
This simulation, labeled Sim.9, had $(20Mpc\,h^{-1})^3$ as box size, $128^3$ as initial grid size and 5 levels of refinement. 
We call initial grid size the number of root grid cells along each axis, which gives us the initial number of cells. 
In our simulation, the refinement is on baryon and dark matter mass and cells are refined when cell mass are 8 times the original cell mass.
We choose to use the Eisenstein and Hu model for initial particles densities and distributions (see ref. \cite{RId134}). 
For star formation and star feedback, we use Global Schmidt Law (see ref. \cite{RId136}). 
As we are interested in the cosmic magnetic fields, we have chosen Dedner magneto-hydrodynamic method implemented in \texttt{ENZO}. 
This method is described in the paper of Dedner et al. (2002)\cite{RId135} (see ref. \cite{RId120} for implementation and test problems). 
In \texttt{ENZO}, only one method to seed magnetic fields is an homogeneous initial magnetic field in all simulation box cells. 
Our equilibrium cooling follows pre-computed tabulated cooling rates and cooling library that is plugged in to \texttt{ENZO}. 
The time integration is carried out with 2nd order Runge-Kutta scheme (see ref. \cite{RId122}). 
Spatial reconstruction employs the piecewise linear method (see ref. \cite{RId123}), and the flux at cell interfaces is computed with the Harten-Lax-van Leer (see ref. \cite{RId124}) approximate Riemann solver. 
This cosmological simulation took 28 hours in 128 cores making the total of 3584 core hours. 
The synthetic universe created had 688 haloes observed but only 443 haloes had star particles. 

For the construction of the synthetic observations, the code \texttt{YT} version 3.1 is chosen which is open source developed and tested by active community. 
The \texttt{YT} code was presented as an analysis code that reads outputs from several simulations codes and creates synthetic observations (see ref. \cite{RId127}) \footnote{http://yt-project.org/}. 
After setting the universe cosmological parameters, we choose the observer redshift to present time $z=0.0$ and a maximum redshift expected to be observed. 
An extended function of the code \texttt{YT} version 3.1 is used to generate this output sequence before the \texttt{ENZO} simulation. 
A new class of functions called (\texttt{haloesLightRay}) is added to the code. 
Those functions had the role of the observation construction using new features added and modifications in old \texttt{YT} functions. 
Each halo found in the available datasets is checked for observation. 
Those halos are identified using implemented method in \texttt{YT}-3.1 and described by Eisentein \& Hu (1998)\cite{RId138}. 
First, our algorithm propagate light of each halo and verify if it reaches the observer in present time. 
Then, the main function proceeds for an observable halo of the current dataset to determine the distance to observer and cosmological redshift. 
It uses a modified version of a \texttt{YT}-3.1 function (\texttt{LightRay}) to collect data from cells that a light ray is passing through. 
Those data include halo particles velocities to compute Doppler effect, gravitational potential in halo's position to compute gravitational redshift and magnetic fields along the light's path to compute magnetically induced redshift. 
At the end of the analysis, a list of observed haloes is presented with their corresponding distance, cosmological, gravitational, Doppler and our new effect redshifts. 
The observation is made to an observer in the center of simulation box. 
This algorithm and preliminary results were firstly presented by {Abdelali \& Mebarki (2015)\cite{RId188}. 
The results presented in the following sections are the most accurate currently achieved.

\begin{table}[tbp]
\centering
\begin{tabular}{|c|c|c|c|c|c|c|c|c|c|c|}
\hline
Halo Id. & $\theta$ & $\phi$ & $z_C$ & $z_D$ & $z_G$ & $z_{MIR}$ & $z_{Obs}$ & $z_{Obs+MIR}$ \\
 & & & $\times 10^{-3}$ & $\times 10^{-4}$ & $\times 10^{-3}$ & &  $\times 10^{-3}$ &  $\times 10^{-3}$ \\
\hline
243 & 0.560 & -2.49 & 3.68 & 16.10 & 1.00 & 1.03 $10^{-7}$ & 4.85 & 4.85\\ 
169 & 0.558 & -2.51 & 3.76 & 0.27 & 0.97 & 1.03 $10^{-4}$ & 4.76 & 4.87\\ 
303 & 0.548 & -2.50 & 3.78 & -1.53 & 0.93 & 1.84 $10^{-6}$ & 4.56 & 4.56\\ 
\hline
\end{tabular}
\caption{\label{tab:TId006} angular position, each redshift effect contribution and observed redshifts for compact halos cluster in Sim.9.}
\end{table}

\section{Results and discussions}
\label{sec:8}

To be able to visualize deviations of observed redshifts, we plot the data on redshift-distance plan, all-sky map and histograms of those effects' contributions for each observed halo. 
The main result is that contributions from different non cosmological redshift effects could be easily confused. 
Those possible confusions are due to the fact that observed redshift is hardly (if not impossible in some cases) broken to its individual effect contributions and that those contributions are in comparable ranges. 
A particularly interesting case is found in all-sky map of sim.9.
We have 3 galaxies labeled (243, 169, 303) with small solid angles between them (see table \ref{tab:TId006}). 
We labeled those haloes by their position in our observed haloes list. 
Those galaxies are close to each other as their distances from observer are around $15Mpc$ (243 at $15.45Mpc$, 169 at $15.82Mpc$, 303 at $15.88Mpc$). 
When analyzing their redshifts contributions, we observe that Doppler effect has different contributions for each galaxy. 
For instance, galaxy 303 has negative Doppler redshift contribution and the two others have positive values. 
This is indicating that they are moving toward each other and that they are interacting gravitationally with each other. 
The gravitational redshift has different but close values. 

\begin{figure}[tbp]
\centering 
\includegraphics[scale=0.5]{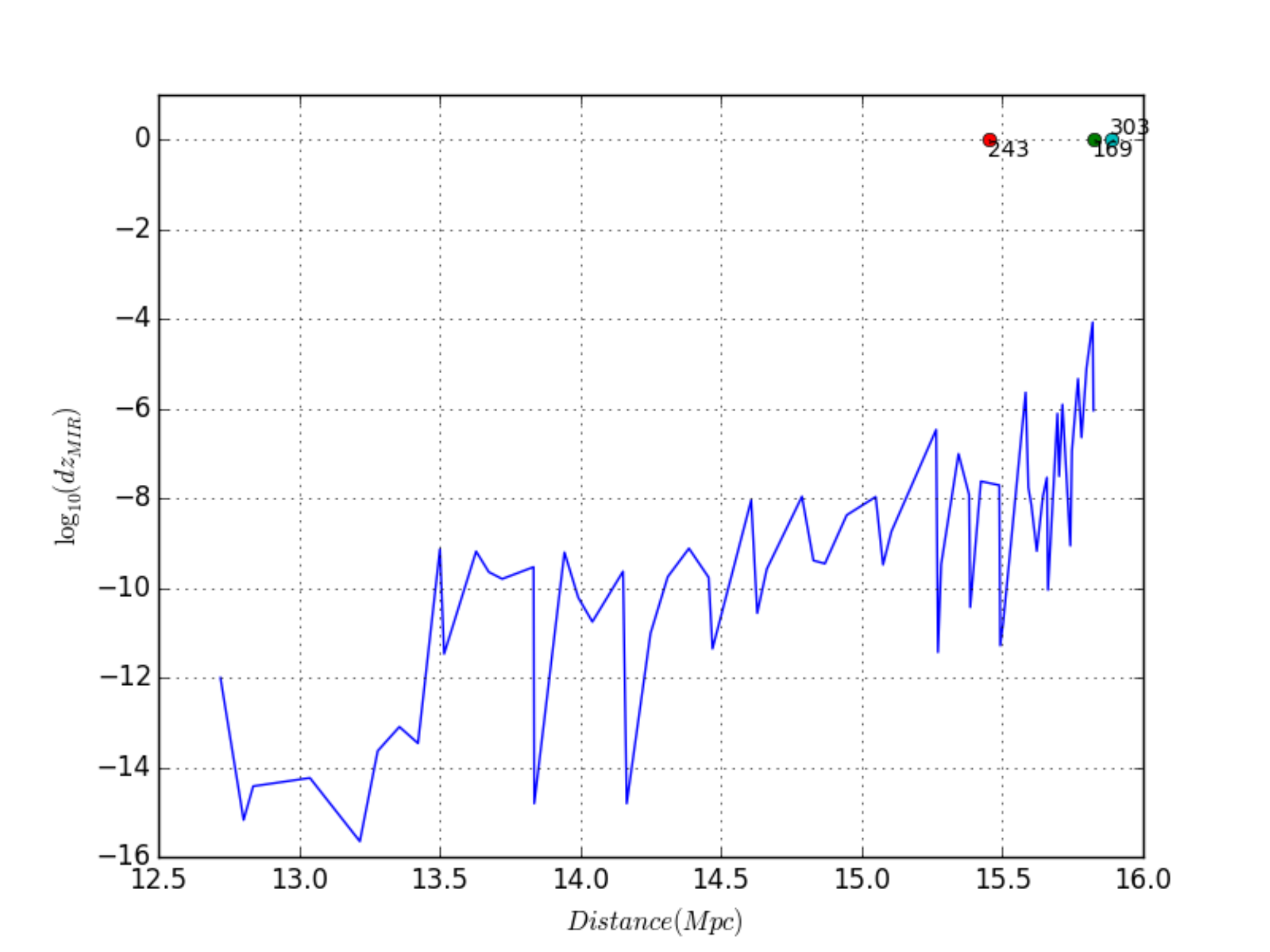}
\caption{\label{fig:FId048} Our redshift effect elementary contributions for each simulation cell to the total redshift of the galaxy 169 of Sim.9 plotted vs. distance to observer and annotated with positions of all galaxies positions.}
\end{figure}

For our magnetically induced redshift, its contributions are very different for each galaxy. 
But, the galaxy 169 has the most significant deviation. 
This is mainly caused by two reasons. 
First, halo 169 is the resulting halo of two haloes merging together and has strong magnetic fields around them given first high contribution from our redshift effect. 
The second reason is that its light will travel near a second region of strong magnetic fields from the foreground galaxy 243 and another merging haloes (see figure \ref{fig:FId048}) getting more amplification. 
When we compare the cosmological redshifts, the differences between galaxies are small and indicating the differences in distances. 
When adding the Doppler effect contribution, the differences are now reflecting mainly the radial velocities that caused the Doppler effect. 
This is giving us a good indication of the kinematics of the compact group. 
Adding the gravitational redshift contribution, as in $z_{Obs}$, affects slightly our perception. 
When the significant contribution from our effect is added, as in $z_{Obs+MIR}$, the differences between observed redshifts are no more indicative of the group members' motion. 
Redshift differences due to cosmological expansion could be seen as difference in Hubble flow and interpreted as velocity differences. 
Along with Doppler contributions, they represent the peculiar motion of group members to each other. 
But, differences from gravitational and especially from our magnetically induced effects could not and should not be interpreted as velocities differences. 
An observer that measures $z_{Obs+MIR}$ and estimate that gravitational contribution is insignificant and do not consider our redshift effect contribution will have an overestimation of velocities differences and an incorrect perception of the group kinematics.

When trying to reproduce formation scenario of SQ, RAX10 searched to recreate observed morphological features and the kinematics of the group. 
Models search to well-reproduce the velocity patterns of the group members existing in previous observational studies (see also ref. \cite{RId149} and ref. \cite{NRId007}). 
Those measures provide only radial velocities and it is not possible to know repartition of those velocities for a given galaxy for recession and proper motion. 
In models proposed by RAX10 (see also ref. \cite{NRId009} and ref. \cite{RId152}), a common recessional velocity is assumed to main group members: NGC 7319, NGC 7320c, NGC 7318b and NGC 7318a. 
The velocity differences represent the geometrical offset along the line-of-sight and a proper motion. 
At the start of each numerical simulation of the quintet, the main galaxy NCG 7319 is assumed to be in the centre of the simulation box with no velocity. 
Other members are given different positions and velocities to reproduce, at the end of the simulation, the relative velocities of the galaxies. 
These velocities' estimations are actually measurements of redshifts interpreted as a combination of cosmological and Doppler redshifts. 
Then, the cosmological part of the redshift corresponds to recession motion and Doppler contribution corresponds to the peculiar motion. 
But, this interpretation is incorrectly neglecting gravitational and do not consider our magnetically induced redshift contributions. 

But, we should consider gravitational and our new effect and the case found in our simulation is supporting this proposal. 
Observed redshifts of SQ members have similar effects contributions to those of the simulated compact group. 
The close positions of the quintet members create high gravitational potential and then significant gravitational redshift contributions. 
The complexity of the quintet induces different gravitational contributions to each member. 
In the paper of Nikiel-Wroczynski et al. (2013)\cite{NRId002}, they presented estimations quintet's magnetic fields from radio observations. 
Their observations had shown significant magnetic fields of $6.4 \pm 1.1 \mu G$ as mean strength in the group area. 
This is supporting a significant contribution of our magnetically induced effect in observed redshifts of quintet's members. 
These fields are also shown to have different strengths and configuration over the quintet and could reach the strength of $11.0 \pm 2.2 \mu G$ between NGC7318A and NGC7319. 
The differences in magnetic fields in quintet regions suggest different contributions from our redshift effect. 
Then, the observed redshift differences of SQ members are the result of several differences of all contributing effects. 
And similar to our simulated compact group, it could not reflect the exact radial velocity differences. 
This fact should be considered with the recent observation indications in the revisited SQ models. 

\section{Conclusions}
\label{sec:9}

Recent studies had shown evidence of non-cosmological and non-Doppler contributions to observed redshifts of extra-galactic objects. 
Gravitational redshifts are responsible of just part of those contributions. 
We introduce a new magnetically induced redshift effect that could be the origin of the rest of redshift bias observed. 
This new effect is caused by a gravitational interaction of photons with cosmic magnetic fields producing gravitational waves and manifested as redshift. 
Customized cosmological simulations were performed to estimate the contributions of our effect to cosmological and astrophysical data. 
In the synthetic observations constructed, a compact group of galaxies similar to SQ is found and gave new indications on SQ kinematics. 
The famous quintet has new observational evidence on its formation history. 
The new indications suggest an older formation with a contribution of NCG 7317 to the successive collisions of group members. 
These indications call for new models and numerical simulations to recreate the formation scenario. 

The velocities of each member represent important parameters to recreate the interaction process. 
Our misinterpretation of redshift measurements of SQ members is creating incorrect estimation of velocities. 
Along with common cosmological redshift and Doppler effect contribution, these redshifts include the gravitational and magnetically induced effect contributions but are not considered by previous studies. 
These velocities are overestimated if the member is moving away from us and underestimated if it moves towards us. 
This is from the fact that our effect is biasing redshift always to higher values. 
Depending on the relative velocities of the galaxies, the final merger would involve all or only part of the members and could occur in more than a Hubble time. 
Until now, different studies suggested that the compact group has recent formation. 
Then, high velocities were needed to explain several morphological features creation. 
For instance, it was believed that the velocity of 7318b should be high to separate it from the rest of the group in a small amount of time. 
Also, the perturbation effects of NCG 7318b is reduced due its high velocity and the short time of their interaction. 

The evidence of an older SQ changes our understanding of the group formation scenario and questions members' believed high velocities. 
It could be an indirect evidence of our overestimation of these velocities affecting deeply our perception and creation of models. 
The role of NCG 7317 in the group history can be misinterpreted. 
An collision between this galaxy and NCG 7318b is already excluded by DCR18 because of the believed high radial velocity of NCG 7318b. 
The new observational evidence with our results should be the ground for future studies aiming to create formation scenarios. 
Our suggestion is that they recreate the redshift patterns of different group members and regions. 
This step will make the new models much accurate and reflecting the reality of the compact group present and historical evolution. 
More results of our simulations and implications of the new redshift will be published in future papers. 

\acknowledgments

Computations described in this work were performed using the publicly-available \texttt{Enzo} code (http://enzo-project.org), which is the product of a collaborative effort of many independent scientists from numerous institutions around the world. 
Their commitment to open science has helped make this work possible. 
Our synthetic observations were done using the publicly-available \texttt{YT} code (http://yt-project.org/). 
The cumulative work of this project's members made our use and development of observations tools more productive. 
This work was granted access to the HPC resources of UCI-UFMC "Unite de Calcul Intensif" of the University Freres Mentouri Constantine 1. 
Authors are every indebted to the Algerian of Education and Research (DGRSDT, ATRST) for the financial support.

\end{document}